\documentclass[12pt]{emulateapj}

\shorttitle{Star formation to $z>1$ in AEGIS field galaxies}
\shortauthors{Noeske et al.}

\begin{document}


\title{Star Formation in AEGIS Field Galaxies since z=1.1 :
 The Dominance of Gradually declining Star Formation,
 and the Main Sequence of Star-Forming Galaxies}


\author{K.G. Noeske\altaffilmark{1}\email{kai@ucolick.org}}
\author{B.J. Weiner\altaffilmark{2}}
\author{S.M. Faber\altaffilmark{1}}
\author{C. Papovich\altaffilmark{2}}
\author{D.C. Koo\altaffilmark{1}}
\author{R.S. Somerville\altaffilmark{3}}
\author{K. Bundy\altaffilmark{4}}
\author{C.J. Conselice\altaffilmark{5}}
\author{J.A. Newman\altaffilmark{6,}\altaffilmark{8}}
\author{D. Schiminovich\altaffilmark{7}}
\author{E. Le Floc'h\altaffilmark{2}}
\author{A.L. Coil\altaffilmark{2,}\altaffilmark{8}}
\author{G.H. Rieke\altaffilmark{2}}
\author{J.M. Lotz\altaffilmark{9}}
\author{J.R. Primack\altaffilmark{10}}
\author{P. Barmby\altaffilmark{11}}
% direct contributors above this line, then alphabetically
\author{M.C. Cooper\altaffilmark{12}}
\author{M. Davis\altaffilmark{12}}
\author{R.S. Ellis\altaffilmark{4}}
\author{G.G. Fazio\altaffilmark{11}}
\author{P. Guhathakurta\altaffilmark{1}}
\author{J. Huang\altaffilmark{11}}
\author{S.A. Kassin\altaffilmark{1}}
\author{D.C. Martin\altaffilmark{4}}
\author{A.C. Phillips\altaffilmark{1}}
\author{R.M. Rich\altaffilmark{13}}
\author{T.A. Small\altaffilmark{4}}
\author{C.N.A. Willmer\altaffilmark{2}}
\author{G. Wilson\altaffilmark{14}}
\altaffiltext{1}{UCO/Lick Observatory, University of California, Santa Cruz, CA 95064}
\altaffiltext{2}{Steward Observatory, University of
Arizona, Tucson, AZ 85721}
\altaffiltext{3}{Max-Planck-Institut f\"ur Astronomie, Koenigstuhl
  17, 69117 Heidelberg, Germany}
\altaffiltext{4}{California Institute of Technology, Pasadena, CA 91125}
\altaffiltext{5}{School of Physics and Astronomy, University of
  Nottingham,  University Park, NG9 2RD, UK}
\altaffiltext{6}{Institute for Nuclear and Particle
Astrophysics, Lawrence Berkeley National Laboratory, Berkeley, CA 94720}
\altaffiltext{7}{Department of Astronomy, Columbia University, New York, NY 10027}
\altaffiltext{8}{Hubble Fellow}
\altaffiltext{9}{Leo Goldberg Fellow, NOAO, Tucson, AZ 85718}
\altaffiltext{10}{Department of Physics, University of California,
  Santa Cruz, CA 95064}
\altaffiltext{11}{Harvard-Smithsonian Center for Astrophysics, 60 Garden St., Cambridge, MA 02138}
\altaffiltext{12}{Department of Astronomy, University of California at
  Berkeley, Berkeley, CA 94720}
\altaffiltext{13}{Department of Physics and Astronomy, UCLA, Los Angeles, CA 90095-1547}
\altaffiltext{14}{Spitzer Science Center, California Institute of
Technology, Pasadena, CA 91125}

\begin{abstract} 
We analyze star formation (SF) as a function of stellar mass
($M_{\star}$) and redshift $z$ in the All Wavelength Extended Groth
Strip International Survey {\it (AEGIS)}. For $2905$ field galaxies,
complete to $10^{10}(10^{10.8})M_{\odot}$ at $z< 0.7(1)$, with Keck
spectroscopic redshifts out to $z=1.1$, we compile SF rates (SFR) from
emission lines, GALEX, and Spitzer MIPS $24\mu$m photometry,
optical-NIR $M_{\star}$ measurements, and HST morphologies.
Galaxies with reliable signs of SF form a distinct {\em ``main
sequence (MS)''}, with a limited range of SFR at a given $M_{\star}$
and $z$ ($1\sigma \la \pm0.3$\,dex), and log(SFR) approximately
proportional to log($M_{\star}$). The range of log(SFR) remains
constant to $z>1$, while the MS as a whole moves to higher SFR as $z$
increases.
The range of SFR along the MS constrains the amplitude of episodic
variations of SF, and the effect of mergers on SFR. Typical galaxies
spend $\sim 67(95)\%$ of their lifetime since $z=1$ within a factor of
$\la 2(4)$ of their average SFR at a given $M_{\star}$ and $z$.  The
dominant mode of the evolution of SF since $z\sim 1$ is apparently a
gradual decline of the average SFR in most individual galaxies, not a
decreasing frequency of starburst episodes, or a decreasing factor by
which SFR are enhanced in starbursts. LIRGs at $z\sim 1$ seem to
mostly reflect the high SFR typical for massive galaxies at that
epoch.
The smooth MS may reflect that the same set of few physical processes
governs star formation prior to additional quenching processes. A
gradual process like gas exhaustion may play a dominant role.

\end{abstract}
%------------------------------------------------------------
%%% MAIN BODY OF TEXT GOES HERE. CONSULT "INSTRUCTIONS FOR AUTHORS USING
%%% LATEX2E MARKUP", SECTIONS 2.3-2.6 FOR HELP WITH EQUATIONS, FIGURES,
%%% AND TABLES.
%------------------------------------------------------------
\section{Introduction}
%------------------------------------------------------------
Deep galaxy surveys have found consistently that the star formation
 rate (SFR)
 per unit stellar mass ($M_{\star}$) depends strongly on
 both $M_{\star}$ and redshift $z$, with the bulk of star formation
 (SF) occurring earlier in massive galaxies than in less massive
 systems (e.g. Guzm\'an et al. 1997, Brinchmann \& Ellis 2000, Juneau
 et al. 2005, Bauer et al. 2005, Bell et al. 2005, P\'erez-Gonzalez
 et
 al. 2005, Feulner et al. 2005, Papovich et al. 2006, Caputi et
 al. 2006, Reddy et al. 2006). High-SFR objects are observed to be
 more abundant at higher $z$; it is often assumed that a part of these
 reflect a greater frequency of merger-driven starburst episodes at
 earlier times. 
However, a comprehensive observational picture of the relationship
between SF and mass to $z\sim 1$, including objects with a wide range
of both masses and SF rates, has been lacking.  
 
 This Letter is
part of a series of papers that study the
 evolution of SFR and
$M_{\star}$ in field galaxies out to $z=1.1$ in the
 All Wavelength
Extended Groth Strip International Survey {\it
 (AEGIS)}.  We combine
SFR measurements from deep Spitzer MIPS $24\mu$m imaging, Keck/DEEP2
spectra, and GALEX UV photometry, allowing us both to recover obscured
SF in IR-luminous galaxies and to include lower-SFR objects not
detected at $24\mu$m. Using optical-NIR
 derived $M_{\star}$
measurements, we analyze the evolution of SFR as a function of
$M_{\star}$ and $z$; we also analyze HST/ACS imaging and rest-frame
colors to support the interpretation of SFR indicators.

 We
adopt a a concordance cosmology ($H_0 =
 70$km\,s$^{-1}$\,Mpc$^{-1}$,
$\Omega_M = 0.3, \Omega_{\Lambda} =
 0.7$). Values of $M_{\star}$ and SFR
are based on a Kroupa (2001) IMF, following recent results by Hopkins
\& Beacom (2006).
%
%-----------------------------------------------------------
\section{Data set}   
%------------------------------------------------------------
Our sample includes all field galaxies with DEEP2 spectroscopic
redshifts $z
 \le 1.1$, in the area where Spitzer MIPS $24\mu$m
photometry and $K$
 band imaging to $22$\,AB\,mag are available; see
Davis et al. 2006 
 (this volume). {\em Stellar masses} were obtained
from SED fits to optical/NIR
 photometry by Bundy et al. (2006);
errors are $<0.3$\,dex, with a mean and rms of 0.1 and 0.05\,dex and
$<4\%$ of errors $>0.2$\,dex. 
Fig. 1 shows data in the $M_{\star}$ range where the sample is $>80
\%$ complete, adopting the completeness analysis by Bundy et
al. (2006; see also Cimatti
 et al. 2006), for a total of 2905
galaxies.  We draw conclusions only where the sample is $>95\%$
complete ($M_{\star}\ge 10(10.8)$\,dex $M_{\odot}$ for $z< 0.7(1)$,
see vertical lines in
 Fig. 1).
For galaxies with robust $24\mu$m detections ($f>60\mu$Jy), SFRs were
derived following Le Floc'h et al. (2005), using Chary \& Elbaz
(2001)
 SED templates; using templates from Dale \& Helou (2002)
yields no
 significant differences. We then add to the $24\mu$m-based
SFR the SFR estimated from DEEP2 emission lines (H$\alpha$, H$\beta$,
or [OII]3727, depending on $z$) with no extinction correction, to
account for SF from unobscured regions.  This approach is similar to
that employed by Bell et al. 2005; utilizing rest-frame UV continuum
SFRs (as they did) in place of emission-line fluxes yields consistent
results. Galaxies below the $24\mu$m detection limit are not dominated
by highly extincted SF; for these, we use extinction-corrected SFR
from emission lines; these can probe to roughly $10\times$ lower SFR
than the
 $24\mu$m data, and are slightly more sensitive at high $z$
and cover a larger area than GALEX data.
 Emission line luminosities
(as calculated in Weiner et al. 2006)
 were transformed to an
H$\alpha$ luminosity using average line ratios
 measured from DEEP2
data (H$\beta$/H$\alpha=0.198$,
 [OII]/H$\alpha=0.69$; Weiner 2006,
private communication), and transformed to SFR
 using the H$\alpha$ calibration
of Kennicutt (1998). The DEEP2
 H$\beta$/H$\alpha$ ratio corresponds
to an extinction of 1.30\,mag at
 H$\alpha$ assuming case B
recombination, which was applied to correct the emission line SFRs. We
use fixed rather than $M_H$-dependent line ratios (a la Weiner et
al.), because these predict extinction corrected SFR slightly in
excess of the $24\mu$m derived SFR for high mass galaxies. Our simple
but robust approach yields results in good agreement
 with SFR
derived from GALEX data, extinction-corrected based on UV spectral
slopes.
 
 For objects with both $f_{24\mu m}<60\mu$Jy and
emission-line $S/N<2$, we estimate a 2-$\sigma$ upper limit on SFR
from the most sensitive emission line available, by adding $2\sigma$
to the measured uncertain SFR, or, for non-detections, to the limit of
$S/N>2$-detectable emission line SFR at the galaxy's redshift.  We
again apply $A_{H\alpha}=1.30$ for extinction corrections, certainly
an overestimate since extinction is lower
 in more weakly SF galaxies
(Hopkins et al. 2001). 
 
 We have performed a suite of tests of
these SF estimates, finding that adopting different SFR tracers
changes results moderately (Noeske et al 2007, in prep.); qualitative
results are unaffected.
 Random errors in our $24\mu$m-based SFRs are
$\la 0.1$\,dex from photometry and $\sim 0.15$\, dex from scatter in
the f($24\mu$m) to L(IR) conversion (see Marcillac et al. 2006), yet
total random errors are expected to be 0.3-0.4\,dex (see Bell et
al. 2005). For extinction-corrected emission line SFRs, random errors
are $\sim
 0.35$\,dex, including scatter about the assumed mean
extinction.
%------------------------------------------------------------
\section{Results}
 %------------------------------------------------------------
\begin{figure*}
\newcommand{\myheight}{7.3cm}
\newcommand{\mywidth}{16cm}%1.18x\myheight
\centerline{\includegraphics[width=\textwidth,clip=]{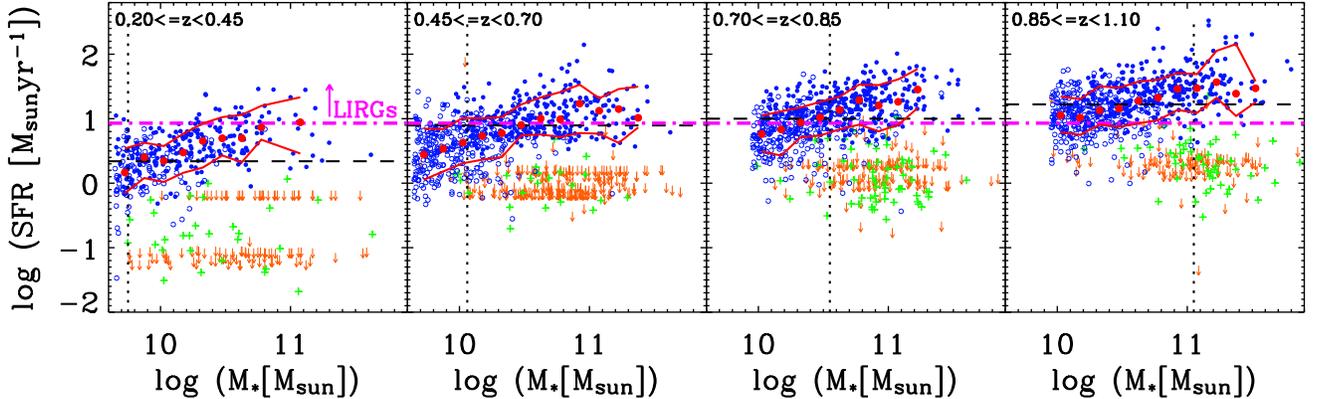}}
\caption{SFR vs $M_{\star}$ for $2905$ galaxies in the EGS, in the
$M_{\star}$ range where the data are $>80\%$ complete; see
\S 2~. The {\it dotted vertical line} marks $>95\%$ completeness. {\it
Filled blue
 circles:} Combined SFR from MIPS 24$\mu$m and DEEP2
emission lines. {\it
 Open blue circles:} No 24$\mu$m detection, blue
$U-B$ colors, SFR from
 extinction-corrected emission lines. {\it
Green crosses:} Same as open blue
 circles, but red $U-B$ colors,
mostly LINER/AGN candidates (\S 3).
 {\it Orange down arrows:}
No robust detection of $f(24\mu$m$)$ or 
 emission lines; conservative SFR
upper limits shown. There is a distinct sequence formed by
 fiducial
SF galaxies (open and filled circles); galaxies with little
 or no SF
lie below this sequence. Red circles show the median
 of log(SFR) in
mass bins of 0.15 dex for main sequence galaxies (blue
 circles). Red
lines include 34\% of the main sequence galaxies above and 34\%
 below
the median of log(SFR), $\pm 1\sigma$ in the case of a normal
distribution. {\it Horizontal 
 black dashed line:} SFR corresponding
to the $24\mu$m 80\% completeness limit at
 the center of each $z$
bin. $24\mu$m-detected galaxies above the {\it magenta dot-dashed
line} are LIRGs (\S 4.2).}
\end{figure*}
%\subsection{A Main Sequence of Star-Forming Galaxies}
\label{ms}

 Fig. 1 shows SFR as a function of $M_{\star}$ in four independent
redshift bins.  The following discussion refers only to the stellar
mass range where the sample is $>95\%$ complete, marked by the
vertical dotted lines in each redshift bin. We identify three
different categories of galaxies:

 (1) The majority of galaxies
show clear signs of SF, either robust
 $24\mu$m detections, or, at
lower $M_{\star}$, blue colors and
 emission lines (blue symbols in
Fig. 1). Quantitative HST morphologies
 (Gini/M20, Lotz et al. 2006;
CAS, Conselice 2003) classify $\la 25\%$ of these galaxies as
 early
types (E,S0,Sa), and $\ga 90\%$ show visual signs of SF such as
 blue
regions and dust lanes. Most of them lie on the ``blue cloud'',
(e.g. Willmer et al. 2006), though some of the massive ones are red,
likely dusty, star-forming galaxies (Bell et al. 2005). This category
(blue symbols in Fig. 1)
 comprises 67(56)\% of the sample at
$z<(>)0.7$ in the $M_{\star}$ range where
 the sample is complete. 

 (2) Clearly separated are galaxies without robust $24\mu$m
($>60\mu$Jy) or emission line ($S/N>2$) detections (orange arrows in
Fig. 1). The upper limits on their SFR are conservatively high
 (\S
2), such that the true separation between the sequence and the
 other
galaxies is likely larger than it appears here.  Almost all ($>95\%$)
of these
 galaxies are on the red sequence, and $\ga 90(80)\%$ at
$z<(>)0.7$
 have early-type quantitative morphologies including
early-type mergers,
 while $\ga 90\%$ at $z>0.7$ have early-type
visual morphologies with
 no hints of current SF.  These galaxies
contribute $29(30)\%$ of the
 sample at $z<(>)0.7$.

(3) Scattered below the star-forming sequence are galaxies with
robust
 emission line detections but no significant $24\mu$m
emission, $5(14)\%$ of the sample at $z<(>)0.7$.  All of these
galaxies (green crosses in Fig. 1) are on the
 red sequence and their
$H\alpha, H\beta$ emission line equivalent
 widths tend to be low
(few \AA ). Yan et al. (2006) and Weiner et
 al. (2006) showed that
the bulk of the line emission in red galaxies
 out to intermediate
redshifts is due to LINER/AGN emission, not SF.
 We find that $75\%$
of those galaxies with [OII] and H$\beta$ detections show LINER-like
line ratios, and $\ga 55(70)\%$ at $z<(>)0.7$ have
 early-type
quantitative and visual morphologies that are typical for local
LINERs (Yan et al. 2006). Line
 emission in these red galaxies thus
appears to be dominated by
 LINERs/AGN, particularly at $z>0.7$ where
they are more
 frequent. Their SFR, derived from emission lines
(Fig. 1), will mostly be overestimated. However, we find visual signs
of SF in the HST images in
 $\la 30\%$ of these galaxies, comparable
to the fraction of non-early
 quantitative morphologies; these may be
dominated by SF.

The star forming galaxies form a distinct sequence of SFR with
$M_{\star}$, which we term the ``main sequence'' (MS). The red lines in
Fig. 1 enclose 34\% of galaxies both above and below
 the median (red
points), and thus indicate the equivalent of $\pm
 1\sigma$ for a
Gaussian distribution. The width of the MS measured in
 this way (the
range in SFR about the median at a given $M_{\star}$)
 is about
$\sigma_{MS} = 0.35$ dex, and seems to remain
 approximately constant
in our sample over the redshift range $0.20 < z
 < 1.1$. Subtracting
a lower limit of non-systematic scatter in SFR
 ($\sim 0.2$\,dex,
\S 2) in quadrature yields an upper limit of
 $\sim 0.3$\,dex on
the {\em intrinsic} scatter, which is still
 broadened by the width
of the $z$ bins, and by additional spread from
 combining different
SFR tracers. Errors in $M_{\star}$ hardly affect
 $\sigma_{MS}$.

The slope of
 the MS is shallower than unity, ${\rm log}(SFR)=(0.67\pm0.08)
{\rm log}(M_{\star})-(6.19\pm0.78)$ for $M_{\star}$ between $10^{10}$ and
$10^{11}M_{\odot}$, and $z=0.2 - 0.7$. There is a trend for the slope
to flatten to higher $z$, but the completeness limits do not allow
a robust quantification.
A further important result is that the normalization of the main
sequence evolves strongly over the redshift range of our sample; the
median SFR at fixed $M_{\star}$ evolves downwards by a factor of 3,
measured at $10^{11}M_{\odot}$, from the our highest (median $z=0.98$)
to our lowest (median $z=0.36$) redshift
 bin. Importantly, it
appears that the {\em whole of the main sequence}
 shifts downwards
with time, rather than just the upper envelope
 decreasing,
which was also reported by a recent GALEX study at $z=0.7$ (Zamojski et
al. 2007). A straightforward interpretation of these observations is
that normal star-forming galaxies possess a limited range of SFR at
a
 given $M_{\star}$ and $z$, which is presumably set by whatever
physical processes regulate SF in quiescent
 disks. Galaxies that are
not on the main sequence, in categories (2)
 and (3) above, are
observed during or after quenching of the SF
 activity, with either
low-level or no current SF, or LINER/AGN
 activity.
 
%------------------------------------------------------------
\section{Discussion}
%------------------------------------------------------------
%\subsection{Amplitude of SF episodes since $z\sim 1$; effect of mergers on SF}

%----------------------------------------
\subsection{Completeness: Is the Main Sequence Real?}
It is obviously crucial to determine whether the ``main sequence''
that we have identified is real, or could be caused by selection
effects or observational biases. We address the following possible
causes of incompleteness or bias in our sample, again restricting
the
 discussion to the $M_{\star}$ range where we claim that the
sample is
 $>95\%$ complete:
 
 {\bf (1)} Could the optically
selected DEEP2 parent sample be missing
 a significant number of
galaxies, or are there galaxies in the DEEP2
 sample that lack a
successful redshift determination because of low
 S/N? {\bf (2)}
Could we be significantly underestimating the SFR in
 galaxies in our
sample due to biases in our SF indicators?
 
 {\bf (1)} The DEEP2
spectroscopic selection ($R_{AB}<24.1$) has been
 shown to be
complete in the $M_{\star}$ ranges indicated in Fig. 1
 (vertical
lines), from comparisons to various surveys with
 spectroscopic and
deep photometric redshifts, including in particular
 the $K$-selected
$K20$ survey, which should be less affected by
 extinction (Willmer
et al. 2006, Bundy et al. 2006, Cimatti et
 al. 2006). For galaxies
that are below our $24\mu$m detection limit,
 we expect the
extinction to be moderate, and would expect these
 galaxies to be
picked up in $K$-selected surveys, but no such
 population is found
to be missed by DEEP2.
 
 More obscured populations can be probed
through the deep Spitzer IRAC
 $3.6\mu$m data in AEGIS, which at a
given redshift are a proxy for
 $M_{\star}$ yet are barely affected
by extinction. We have compared
 the distribution of $f$($24\mu$m) at
a given $f$($3.6\mu$m) and $z$ in
 the DEEP2 $R_{AB}$-selected sample
and an IRAC $f$($3.6\mu$m)-selected
 sample with IRAC-based
photometric redshifts. We find no evidence that
 DEEP2 misses a
significant population of heavily obscured,
 star-forming galaxies at
$z\la 1$, which could populate the area
 above the upper boundary of
the MS. This agrees with the results of
 Houck et al. (2005) and
Weedman et al. (2006) in the large area NDWFS,
 which indicate that
such missed $f$($24\mu$m)-bright, optically faint
 galaxies at $z<1$
would contribute $< 1\%$ of our sample.
 
 {\bf (2)} The $24\mu$m
completeness limit (horizontal black dashed
 line in Fig. 1)
intersects the main sequence in each redshift bin.  As
 discussed in
\S \ref{ms}, most galaxies below the MS are red, early
 type, non-SF,
and/or LINER/AGN dominated (shown as orange arrows and green
crosses).
 However, a fraction show spiral/late-type morphologies or
visual signs of possible SF (\S 3). In principle, these red galaxies
could have dust-obscured SF, unrecovered by emission lines,
 yet lie
below the $24\mu$m detection limit. Their true SFRs could then be
anywhere up to the $24\mu$m limit, in which case they may not be a
distinct population, but rather a downward continuation of the
MS. If
 this were the case, these galaxies would make up $\la
10(20)\%$ of the
 MS at $z<(>)0.7$. We can constrain the maximal
effect of missed,
 dust-obscured SF in these galaxies on the
1$\sigma$ range of SFR along
 the MS by including in the calculation
of $\sigma_{MS}$ all red, $24\mu$m-undetected galaxies with
spiral/late-type morphologies, and H$\alpha$,H$\beta$, and/or [OII]
line emission down to
 spurious detections (i.e., 100\% error in
EW). For the extremes of either only the emission line SFR or the
maximal SFR, corresponding to the $24\mu$m limit, the measured width
of the MS
 increases by $\sim 0.05$\,dex or not at all, respectively.

 Thus we argue that the relatively sharp upper limit of the MS is
real,
 as our selection does not miss obscured sources with high
SFR. The
 sharpness of the lower limit is more uncertain with our
current data,
 but we find that only a small fraction of galaxies
that we placed
 below the MS could have underestimated SF rates
which
 would ``blur out'' the lower edge of the MS.  Very deep
24$\mu$ data
 at $z\sim 1$ from GOODS (Elbaz 2006, private
communication)
 unambiguously confirm this result, particularly a
well-defined lower
 boundary to the sequence.

\subsection{Constraints on Episodic Star Formation}

 Studies based on local samples (Brinchmann et al. 2004, Salim et
al. 2005 for SDSS; Lee 2006) have illustrated a relationship between
SFR and $M_{\star}$, and have identified two populations: galaxies
on
 a star-forming sequence, and ``quenched'' galaxies with little or
no
 detectable SF. At higher $z$, previous studies (see
 \S 1) had
merely described an upper envelope of SFR in the
 SFR- $M_{\star}$
diagrams. We have employed a variety of SFR tracers
 and other
evidence from AEGIS to show that the SF sequence persists
 out to
$z\sim 1$, with a similar dispersion in log(SFR) at fixed
$M_{\star}$ but with a decrease in normalization of a factor of 3
from
 $z=0.98$ to $z= 0.36$, measured at
$M_{\star}=10^{11}M_{\odot}$. The {\em global} star
 formation rate
density has also decreased by a factor of 3 over this
 same interval
(Hopkins et al. 2004, Eq. 3).  One possible physical explanation for
this
 decline is a decreasing contribution from starbursts in
gas-rich
 galaxy mergers.  However, if this were the dominant factor
causing the
 decline, we would expect to see the upper envelope of
the main
 sequence move downwards with time, with the region
populated by
 ``normal'' galaxies maintaining the same
normalization. This is
 contrary to what we see in AEGIS: the region
of the SFR-$M_{\star}$ space
 populated by main sequence galaxies at
$z\sim0$ (Brinchmann et al. 2004) is nearly empty at $z\sim0.7$--1,
though these galaxies should be detectable in
 our survey.

We can use our observed MS to quantitatively constrain the duty
cycle
 of episodic variations of SFR around an average level. We
adopt the
 densely populated peak of the SFR distribution (the
median) as this
 baseline level. The $1(2)\sigma$ ranges about the
median of log(SFR),
 $\pm 0.3(0.6)$\,dex, include 68(95)\% of the
galaxies. We can hence
 infer that SFR variations exceeding $\pm
0.3(0.6)$\,dex, factors of
 2(4), have duty cycles $<32(5)\%$. These
correspond to total times of
 $<2.5(0.4)$\,Gyr since $z=1$. The
amplitude of these variations ($\la
 \times 4$) is consistent with
gas-poor or minor mergers, rather than
 the peak SFR of gas-rich
major mergers (Springel 2000, Cox et al. 2006). Excursions in
 SFR
$>5\times$ above the median are rare, $\sim 1\%$, consistent with
galaxies spending $\la 100$\, Myr in such strong burst episodes
since
 $z=1$. Of course these arguments are only valid in a
statistical
 sense: a fraction of galaxies could have a lower average
level of SF
 and undergo larger excursions, but at the expense of
reducing the
 allowed range of SFR for the remainder of the
population.
 
 Previous studies have found that $\la 30$\% of SF at
$z\sim 0.7$
 occurs in morphologically disturbed galaxies (Wolf et
al. 2005; Bell
 et al. 2005) or close pairs (Lin et
al. 2006). Semi-analytic models
 predict that about 5\% of the SF at
$z\sim0.7$ is due to
 major mergers, with the contribution due to
minor mergers being more
 uncertain, but ranging from $\sim 11$--45
\% (Somerville et al. 2001;
 Wolf et al. 2005). These direct
constraints and theoretical
 expectations are consistent with the
conclusions that we have drawn
 here from the SFR-$M_{\star}$
distributions.  
A related comment pertains to the nature of Luminous Infrared Galaxies
(LIRGs, $L(8-1000\mu )>10^{11}L_{\odot}$). LIRGs at $z\sim 0$ are
rare, mostly interacting galaxies (Sanders \& Mirabel 1996) with
strong starbursts (SFR $\ga 5\times$ above those of typical
spirals). At $z\sim 1$, LIRGs seem to mostly represent the high level
of SFR in almost all massive SF galaxies, rather then extreme
starbursts (Fig. 1)~.
%A related comment pertains to the nature of Luminous Infrared galaxies
%(LIRGs, $L(8-1000\mu )>10^{11}L_{\odot}$) and the physical cause of
%``downsizing'', or the decreasing contribution of high-luminosity
%relative to low-luminosity galaxies to the global SFR density (e.g. Le
%Floc'h et al. 2005). LIRGs at intermediate $z$ were commonly assumed
%to represent brief starburst episodes (see, e.g., Wang 2006, Hammer et
%al. 2005 and references therein), possibly merger-driven, in analog to
%local LIRGs. However, studies of LIRGs at $z\ga 0.7$ (Bell et
%al. 2005, Melbourne et al. 2005, Lotz et al. 2006) have found mostly
%late-type, non-merger morphologies. The work presented here suggests
%that the rapid decline of the LIRG population and their contribution
%to the global SFR density is a result of the shifting normalization of
%the main sequence, rather than a decline in the merger rate.

In summary, we suggest a picture in which we are witnessing a gradual
decline in the SFR of most galaxies since $z\sim1$, accompanied by
rapid quenching in a fraction of (massive) galaxies. Presumably the
regularity and constant dispersion of the main sequence out to $z\sim
1$ means that the same physics that regulates SF in local disk
galaxies is operating, indicating significant evolution either in the
gas supply or SF efficiencies over this interval.

%Assuming that $SFR \propto m_{gas}/\tau_{dyn}$, as
%in local disks (Kennicutt et al. 1998), where $m_{gas}$ is the
%mass of cold gas available for SF and $\tau_{dyn}$ is
%the orbital period of the disk, our finding that the normalization of
%the main sequence has decreased by a factor of $\sim 5$ since $z\sim
%1$ then implies either a similar decrease in the average gas fractions
%of galaxies above our mass completeness limit, or a corresponding
%increase in the dynamical time at fixed stellar mass, or some
%combination of the two. 
In the accompanying Letter (Noeske et al. 2007, this volume), we show
that the slope and evolution of the MS can be understood as gradual
gas exhaustion in a model in which galaxy age and SF timescale are a
function of galaxy mass, and the dispersion of the MS is interpreted
as resulting from a spread in age and SF timescales at a given mass.

\acknowledgements See the survey summary paper (Davis et al. 2006,
  this volume) for full acknowledgments.  This work is based on
  observations with the W.M. Keck Telescope, the Hubble Space
  Telescope, the Galaxy Evolution Explorer, the Canada France Hawaii
  Telescope, and the Palomar Observatory, and was supported by NASA
  and NSF grants. We wish to recognize the cultural role that the
  summit of Mauna Kea has within the Hawaiian community.  This work is
  based in part on observations made with the {\it Spitzer Space
  Telescope\/}, which is operated by the Jet Propulsion Laboratory,
  California Institute of Technology under a contract with NASA.
  Support for this work was provided by NASA through contract numbers
  1256790, 960785, and 1255094 issued by JPL/Caltech. KGN acknowledges 
  support from the Aspen Center for Physics. We wish to thank the referee 
  for very valuable comments, and D. Elbaz and J. Lee
  for helpful discussions.

%%------------------------------------------------------------
%\acknowledgements See the survey summary paper (Davis et al. 2006,
%this volume) for full acknowledgments.  This work is based on
%observations with the Spitzer Space Telescope, the W.M. Keck
%Telescope, the Hubble Space Telescope, the Galaxy Evolution Explorer,
%the Canada France Hawaii Telescope, and the Palomar Observatory, and
%was supported by NASA and NSF grants. KGN wishes to thank Dr. J. Lee
%for enlightening discussions and sharing her results on local
%star-forming galaxies. We wish to recognize the cultural role that the
%summit of Mauna Kea has within the Hawaiian community. It is a
%privilege for us to carry out observations from this mountain.
%%------------------------------------------------------------

%\clearpage

\end{document}